\documentclass[reprint,amsmath,amssymb,pra,showpacs]{revtex4-1}

\usepackage{graphicx}
\usepackage{dcolumn}
\usepackage{bm}

\begin{document}

\title{Performance predictions for a laser intensified thermal beam for use in high resolution Focused Ion Beam instruments}

\author{S H W Wouters}
\author{G ten Haaf}
\author{R P M J W Notermans}
 \altaffiliation{Present address: LaserLaB, Department of Physics and Astronomy, VU University Amsterdam, De Boelelaan 1081, 1081 HV Amsterdam, Netherlands}
\author{N Debernardi}
 \altaffiliation{Present address: Equipment for Additive Manufacturing, Netherlands Organization for Applied Scientific Research TNO, De Rondom 1, 5612 AP Eindhoven, the Netherlands}
\author{P H A Mutsaers}
\author{O J Luiten}
\author{E J D Vredenbregt}
 \email{Corresponding author: e.j.d.vredenbregt@tue.nl}
\affiliation{Department of Applied Physics, Eindhoven University of Technology, P.O. Box 513, 5600 MB Eindhoven, the Netherlands}

\date{\today}

\begin{abstract}
Photo-ionization of a laser-cooled and compressed atomic beam from a high-flux thermal source can be used to create a high-brightness ion beam for use in Focus Ion Beam (FIB) instruments. Here we show using calculations and Doppler cooling simulations that an atomic rubidium beam with a brightness of $2.1\times 10^7$ A/(m${}^2$\,sr\,eV) at a current of 1 nA can be created using a compact 5 cm long 2D magneto-optical compressor which is more than an order of magnitude better than the current state of the art Liquid Metal Ion Source.
\end{abstract}  

\maketitle


\section{Introduction}
The Focused Ion Beam (FIB) is a valuable tool in the semiconductor industry since it enables imaging and modification of structures on the nanometer size scale \cite{RaffaFIB}. The most important property of a FIB, the spotsize versus current curve, is largely determined by the transverse reduced brightness (hereafter abbreviated as the brightness) and longitudinal energy spread of its ion source. The current state-of-the art for modification of structures is the Liquid Metal Ion Source (LMIS) which creates a gallium ion beam with a brightness of $10^6$ A/(m${}^2$\,sr\,eV) and a longitudinal energy spread of 4.5\,eV \cite{OrloffFIB,Hagen2008}. Note that the Gas Field Ionization Source (GFIS) is a promising alternative for ion beam imaging with a brightness of more than $10^9$ A/(m${}^2$\,sr\,eV) \cite{Ward2006} and a longitudinal energy spread of less than 1\,eV \cite{Ward2006} allowing sub-nanometer resolution. Due to the lower sputter yield and subsurface damage of the helium \cite{Ward2006} and neon \cite{Livengood2011} based GFIS, this apparatus is less suitable for sample modification than the LMIS \cite{Tan2010}. In order to create a FIB with the possibility of high-resolution sample manipulation a heavy ion based source is required with a smaller longitudinal energy spread and at least equal brightness than the LMIS. Furthermore, currents up to 1 nA should be possible and a compact source is preferred.   

Several research groups are working on reaching these goals. The Nano Aperture Ion Source \cite{KruitNAIS} for example aims at creating an ion beam by electron impact ionization of a high density gas. A different idea is to use laser-cooled atoms as source for cold ions as was proposed by Freinkman et al. \cite{FreinkmanIdea}. Apart from promising high-brightness ion beams, laser cooling can be applied to a variety of atomic species ranging from the alkali and alkaline metals, several transition and rare-earth metals and some p-block materials which would open up new possibilities for FIB users. The Ultra Cold Ion Source (UCIS) \cite{vdGeer2007} and Magneto-Optical Trap Ion Source (MOTIS) \cite{KnuffmanMOTIS} both use laser cooling and trapping in 3D followed by in-field photo-ionization to create ion bunches or beams. Although longitudinal energy spreads down to 20 meV have been demonstrated \cite{ReijndersUCIS} the target brightness could not be achieved with these sources \cite{Debernardi2012a}. This is because refilling the ionization region in the MOT is slow due to the low diffusion rate caused by the small velocity of the atoms. To circumvent this problem one could increase the loading rate by addition of a 2D MOT or Low-Velocity Intense Source (LVIS) \cite{LuLVIS} to the system. Omitting the 3D MOT completely is a more direct way of reaching the same goal. Knuffman et al. \cite{Knuffman2DMOT} have reported on a FIB based on a vapour cell 2D+ MOT combined with two-step photo-ionization with an inferred brightness of ${10^7}$ A/(m${}^2$\,sr\,eV) at a current of several picoamperes. Cooling and compressing atoms originating from a thermal source (such as a Knudsen cell) allows for the creation of beams with equal brightness but even higher currents since the availability of the very large flux from such a source allows the selection of the best part of the laser cooled atom beam and thus keeping high brightness up to currents of the target 1 nA. The FIB source considered here, as the one of Kime et al. \cite{KimeComparat}, therefore uses a Knudsen cell as atomic source. In our case, the Knudsen cell is connected to a heated collimation tube \cite{OlanderTube,ScholtenTube} which increases the lifetime of the atomic reservoir by more than an order of magnitude and alters the velocity distribution such that even more atoms can be captured in the magneto-optical compressor (MOC).  

\begin{figure}[t]
	\includegraphics[width=1.0\linewidth]{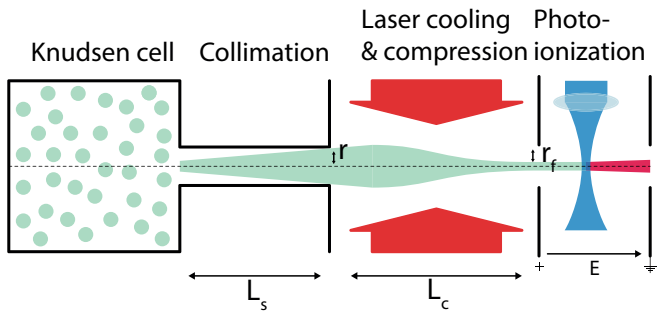}
	\caption{\label{fig:beamline}Schematic drawing of the Atomic Beam Laser-cooled Ion Source in the xz-plane. From the left to the right: an atomic beam is formed from a Knudsen cell connected to a collimating tube with a radius $r$ and length $L_s$. These atoms are then laser-cooled and compressed over a distance $L_c$ to a final radius of $r_f$. At the right the atomic beam is photo-ionised in an electric field $E$ to create the ion beam.}
\end{figure}

Fig. \ref{fig:beamline} shows a schematic representation of the proposed source. The atomic beam is formed from a Knudsen cell with an aperture radius $r$ connected to a collimating tube with equal radius $r$ and length $L_s$. After leaving the tube, the atoms are introduced into a magneto-optical compressor with length $L_c$ which cools the atoms in the transverse direction to a mK temperature and compresses it to a beam with a radius $r_f$. Immediately after leaving the compressor, the atoms are photo-ionized in a two-step process by two crossed laser beams. At the intersection of these beams, ions are formed which are accelerated by the electric field $E$ between two accelerator plates. The resulting ions are focused by an (electrostatic) focusing column.

This design can be used for a broad range of atomic species (except for the meta-stable ones) although the detailed design of the Knudsen cell and collimating tube will look different for elements with a high melting point. As a specific example, rubidium will be used which has a rich history in laser cooling and is, due to its higher mass, expected to be more effective for ion beam milling than gallium. All relevant constants for the two stable isotopes of rubidium are given in Tab. \ref{tab:Constants}.  

\begin{table}[t]
	\small
	\caption{\label{tab:Constants}Atomic constants used in the calculations and the simulations. All data is taken from \cite{SteckRb} except when indicated otherwise. Note that the values for $p^*$ and $T^*$ are only valid for temperatures between 312 K and 550 K.}
	\begin{tabular}{l|r|r|r}
		\hline
		Parameter (unit) & Symbol & ${}^{85}$Rb & ${}^{87}$Rb \\
		\hline
		Abundance (\%) & $ab$ & 72.2 & 27.8 \\
		Mass (amu) & $m$ & 84.91 & 86.91 \\
		Nuclear spin quantum number (-) & $I$ & 5/2 & 3/2 \\
		Pressure constant ($10^9$ Pa) & $p^*$ & $2.05$ & $2.05$ \\
		Temperature constant ($10^3$ K) & $T^*$ & $9.30$ & $9.30$ \\
		Van-der-Waals radius (pm) \cite{CRC_Radii} & $r_{vdw}$ & 303 & 303 \\
		Natural linewidth (MHz) & $\Gamma/2 \pi$ & 6.07 & 6.07 \\
		Cooling wavelength (nm) & $\lambda$ & 780 & 780 \\
		Saturation intensity (W/m${}^2$) & $I_{sat}$ & 16.7 & 16.7 \\
		Doppler temperature limit ($\mu$K) & $T_D$ & 146 & 146 \\
		\hline
	\end{tabular}
\end{table}

This paper discusses calculations and simulations of the proposed source that show that the desired equivalent atomic beam brightness and current can be achieved using realistic parameters for the atomic source and the MOC. First, an analytical model is set up to predict the performance of the system. The analytical model is verified using Doppler cooling simulations which are expanded to include the effects of the real atomic structure of rubidium and the distributions that can be expected from a real thermal source. Finally, a set of parameters is searched which allows the laser cooler to achieve the desired brightness and current within only a compact MOC. 

\section{Analytical model}
An analytical model is set up based on standard theoretical treatment of laser cooling as formulated by Metcalf and Van der Straten \cite{MetcalfBook} to give a understanding of the relevant parameters of the source. In short, the brightness is calculated of a skimmed thermal beam from a Knudsen cell that is cooled to the Doppler temperature and compressed to such a radius that the kinetic energy of the atoms equals their spring energy. The actual atomic structure of rubidium is simplified into a F=0 ground state and F'=1 exited state. Furthermore, the 2D problem is treated as quasi-2D by assuming the forces in the two transverse directions are independent and can be summed.   

The flux originating from a Knudsen cell at an atomic density $n(T_s)$ with a circular aperture of radius $r$ is given by \cite{ScolesBook}
\begin{equation}
	\label{eq:fluxKn}
	\Phi_{tot}={}^1/_4~n(T_s)~\pi r^2~\langle v \rangle ,
\end{equation}
with $\langle v \rangle=\sqrt{8 k_B T_s/ \pi m}$ the average velocity for a gas in thermal equilibrium in the Knudsen cell at temperature $T_s$, $k_B$ Boltzmann's constant and $m$ the mass of the atom. For alkali metal vapours above the melting temperature, the atomic density can be approximated by \cite{CRC_Vapour}
\begin{equation}
	n(T_s) =\frac{p^*~ exp(-T^*/T_s)}{k_B T_s} ,
\end{equation}
with $p^*$ and $T^*$ constants for the specific atomic element as given in Tab. \ref{tab:Constants}. 

Skimming the Knudsen cell by using an aperture with radius $r$ at a distance $L_s$ provides a flux through this aperture of
\begin{equation}
	\label{eq:fluxApertured}
 	\Phi_{skimmed} = \Phi_{tot}~\theta^2 ,
\end{equation}
with $\theta=r/L_s\ll 1$ the opening angle. For the brightness is a figure of merit describing ion, not atomic, beams, we assume the atomic beam can be ionized with a 100\% efficiency resulting in a current of 
\begin{equation}
	\label{eq:currentApertured}
	I=e\Phi_{skimmed},
\end{equation}
with $e$ the electron charge. From the current and transverse temperature of the thermal source, the transverse reduced brightness of an ion beam with the same properties as the initial atomic beam  can now be calculated using the brightness of a thermal emitter \cite{LuitenBrightness}: 
\begin{equation}
	\label{eq:BrInitial}
	B_r^i = \frac{e J}{\pi k_B T_t} = \frac{e^2 \Phi_{tot}}{\pi^2 r^2~ k_B T_s},
\end{equation}
where $J=I / \pi r^2$ the current density and $T_t=T_s \theta^2$ the effective transverse temperature of the atoms leaving the skimmed Knudsen cell.

Filling in the temperature from Tab. \ref{tab:Analytical}, which will be explained further on, results in a brightness of $1.3 \times 10^3$ A/(m${}^2$\,sr\,eV) which is four orders of magnitude lower than desired. Laser cooling can decrease the transverse temperature of the atoms to the Doppler temperature \cite{FootBook}, increasing the brightness by a factor $T_s \theta^2/T_D$. Using the opening angle from Tab. \ref{tab:Analytical} results in a factor $10^2$ increase and a brightness of $1.5\times 10^4$ A/(m${}^2$\,sr\,eV) which is still insufficient for use in a FIB.

Freinkmans idea \cite{FreinkmanIdea} for a laser-cooled FIB can be improved by applying compression to the beam using a magnetic field gradient and $\sigma^+ - \sigma^-$ polarised laser beams. Assuming equilibrium has been reached once the spring and kinetic energy of the particles are equal: $k_B T_D = r_f^2 \kappa$ with $\kappa$ the spring constant, the atomic beam can be compressed to an area of $A_f=\pi r_f^2 = \pi k_B T_D / \kappa$.

With the Doppler temperature and spring constant given by \cite{FootBook}
\begin{equation}
	\label{eq:TDoppler}
	T_D = - \frac{\hbar \Gamma}{4} \frac{\Gamma}{\delta} \left(1+s_0 +(2 \delta / \Gamma)^2 \right) ,
\end{equation}
\begin{equation}
	\label{eq:Kappa}
	\kappa =-\frac{16 \pi \mu_B \nabla B s_0}{\lambda} \frac{\delta / \Gamma}{\left(1+s_0+(2\delta /\Gamma)^2\right)^2} , 
\end{equation}
with $\hbar$ the reduced Planck constant and $\mu_B$ the Bohr magneton, the final brightness of the compressed beam now reads

\begin{equation}
	\label{eq:BrFinal}
	\begin{split}
	B_r^f=B_r^i\frac{T_s \theta^2}{T_D}\frac{\pi r^2}{Af} = B_r^i\frac{T_s \theta^2}{T_D}\frac{r^2 \kappa}{k_B T_D} \\
	  \propto n(T_s) ~{r}^2 ~\theta^2 \frac{\nabla B ~\delta^3 ~s_0}{\left(1+s_0+(2 \delta / \Gamma)^2\right)^4} .
	\end{split} 
\end{equation}
This result shows that the detuning and saturation parameter have an optimum at $\delta=-\Gamma/2$ and $s_0=I/I_{sat}=2/3$, but increasing the source radius $r$, the source opening angle $\theta$ or the magnetic field gradient $\nabla B$ will result in unbounded growth of the brightness. This is due to the assumption that all particles coming from the Knudsen cell within the angle $\theta$ can be cooled and compressed which is only true for an infinitely long MOC.

The following part describes how the brightness for a MOC with finite length can be found by placing limits on $r$, $\theta$ and $\nabla B$. The limit on $r$ is based on the finite displacement an atom experiences by the magneto-optical force during the average interaction time  $t = L_c / \langle v \rangle$. The limits on $\theta$ and $\nabla B$ are based on the assumption that atoms can only be cooled and compressed if their velocity smaller is than the capture velocity and their position is within the capture radius. 

The average acceleration that an atom experiences due to the magneto optical compression force $F_{MOC}$ \cite{MetcalfBook} while being compressed from the capture range $x_c$ to the axis is assumed to be
\begin{equation}
	\label{eq:Aavg}
	a_{avg} = \lim_{x~\to~x_c,v_x~\to~0} \frac{F_{MOC}(x,v_x)}{2 m}.
\end{equation}
The factor $1/2$ takes into account that the force decreases to zero when the atom is slowed down and pushed to the axis \cite{FootBook}. Since this acceleration is finite, as is the transit time of the atoms, the maximum displacement that can be achieved is given by $r_{max} = \frac{1}{2}a_{avg}~t^2$. The initial aperture radius is now set to this maximum position, $r=r_{max}$, to ensure all particles can be compressed. 

Cooling and compression is only efficient when the force on the atom is in the linear regime: $F_{MOC} \approx - \kappa x + \alpha v_x$,  which holds for particles with a transverse velocity lower than the capture velocity $v_c$ and a position smaller than the capture range $r_c$. The capture velocity is typically $v_c = \lambda \Gamma / 4 \pi$ which means that in the paraxial approach the opening angle should be constrained by $\theta = v_c/\langle v \rangle = \lambda \Gamma / (4 \pi \langle v \rangle)$. The magnetic field gradient is constrained by setting the capture range equal to the aperture size ($r=x_c=\hbar \Gamma / (2 \mu_B \nabla B)$, resulting in $\nabla B = \hbar \Gamma / (2 \mu_B r)$. 
The values of these constraints are calculated and listed in Tab. \ref{tab:Analytical}. Using these values, the compression increases the brightness by a factor $5\times 10^2$ to $7.3\times 10^7$ A/(m${}^2$\,sr\,eV) (as calculated with Eq. \ref{eq:BrFinal}) which is more than an order of magnitude higher than that of the LMIS. The flux (as calculated with Eq. \ref{eq:fluxApertured}) is larger than the desired value of $6.2 \times 10^9$ s${}^{-1}$ (1 nA). 

The last free parameter to be discussed is the source temperature $T_s$ on which the brightness is exponentially dependent and can thus be used to make a substantial improvement. This parameter is however also bound to a maximum value since Eq. \ref{eq:fluxApertured} and Eq. \ref{eq:BrInitial} are only valid if no inter-atomic collisions occur between the Knudsen cell and the skimming aperture. An increase in temperature leads to a higher pressure resulting in a shorter mean-free-path $\lambda_{mfp}$ and thus to more inter-atomic collisions in the space between the Knudsen cell and the skimming aperture. The Knudsen number \cite{ScolesBook} describes the importance of collisions by the ratio between the mean-free-path and relevant dimension $x$: 
\begin{equation}
	K_{n,x}=\frac{\lambda_{mfp}}{x} =  \frac{1}{x} \left(4 \sqrt{2} \pi~{r_{vdw}}^2~n(T_s) \right)^{-1}
\end{equation}
where $r_{vdw}$ is the Van der Waals radius of the atom as given by Tab. \ref{tab:Constants}. A Knudsen number higher than unity indicates that the inter-atomic collisions can be neglected. For the analytical model to be valid, the Knudsen number related to the distance between the Knudsen cell and the skimming aperture $L_s=r/\theta$ to be higher than one is thus required. The results in Tab. \ref{tab:Analytical} shows that this is indeed the case for the chosen temperature.

\begin{table}[t]
	\small 
	\caption{\label{tab:Analytical}Parameters given by the analytical model. The top part lists the chosen parameters, the middle part the constraints and the bottom the results.}
	\begin{tabular}{l|r|r}
		\hline
		Parameter (unit) & Symbol & Value \\
		\hline
		Source temperature (K) & $T_s$ & $383$ \\
		Cooling and compression length (m) & $L_c$ & $0.05$ \\
		Saturation parameter (-) & $s_0$ & $0.67$ \\
		Detuning ($\Gamma$) & $\delta$ & $-0.5$ \\
		\hline
		Initial aperture radius (mm) & $r$ & $0.21$ \\
		Opening angle (mrad) & $\theta$ & $7.7$ \\
		Skimmer position or tube length (mm) & $L_s$ & $27.3$ \\
		Magnetic gradient (T/m) & $\nabla B$ & $1.0$ \\
		\hline \hline
		Brightness (A/m$^{2}$ sr eV) & $B_r$ & $7.3\times 10^7$ \\
		Flux (s${}^{-1})$ & $\Phi_{skimmed}$ & $7.0\times 10^9$ \\
		Knudsen-length number (-) & $K_{n,L_s}$ & $2.0$ \\
		\hline
	\end{tabular}
\end{table}

\section{Simulations}
In the previous section an analytical model was introduced that allows to make initial predictions about the brightness that can be expected from a compact MOC. The model is however a simplification of the actual cooling and compression mechanism and may not yield realistic results for the performance of a laser-cooled collimated Knudsen cell. Doppler-cooling simulations \cite{VredenbregtCOOL} allow for a more accurate description of the laser-cooling and compression process. The simulation software traces atoms trough a light field while taking the effect of each absorption-emission cycle into account. Mind that the atoms do not interact with each other and thus no collisions are taken into account. In order to allow for quasi-2D results, pairs of half-atoms are traced trough a 1D light and magnetic field. The half-atoms in such a pair have identical longitudinal velocity but different transverse positions and velocities and thus represent the x- and y-coordinates of one full atom. In the post-processing these half-atoms are combined and the brightness and flux figures can be calculated for collections of these atoms. Depending on the parameters, $10^5$ to $10^7$ atoms need to be traced in order to have good statistics. The code has also undergone modifications to include the starting distributions for both a skimmed and a collimated Knudsen cell. These distributions were found by a geometrical calculation and Monte-Carlo simulations under the assumption that the atoms only interact with the wall of the collimation tube (thus $K_n \gg 1$).

Before using the simulations for scenarios surpassing the analytical model, it is checked against this model. Then the effects of including the actual level structure of rubidium are investigated, followed by the collimation tube on the Knudsen cell. Finally, the parameters are varied to find the optimal performance of the MOC.   

\subsection{Verification}
The simulation is checked against the analytical model by looking at the scaling of the brightness with the laser detuning. In order to do so, a hypothetical atom is assumed which has all the physical properties of the ${}^{85}$Rb isotope except for the ground-state having the total angular momentum quantum number F=0 and the excited-state F'=1. This is done to mimic the two-level atom that is used in the analytical model. For the thermal source the atom longitudinal velocity is taken equal to the average velocity, the position is uniformly distributed over a circle with radius $r$ and the angle uniformly distributed between $\theta$ and $-\theta$. In the following analysis the brightness of the 10\% fraction of the atoms which are closest to the axis at the end of the MOC is reported because this provides a good indication of the center brightness. 

\begin{figure}[t]
    \centering
        \includegraphics[width=1.0\linewidth]{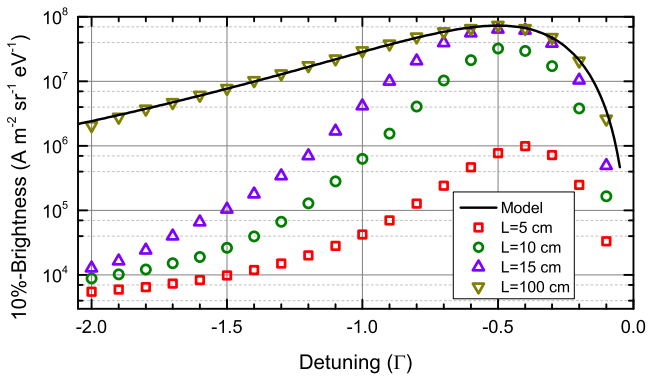}
    \caption{\label{fig:comp_detu} 10\% brightness $B_r^{10\%}$ plotted on a logarithmic scale as a function of the detuning, for different cooling and compression stage lengths $L$. As $L$ increases, the beam is cooled and compressed more into equilibrium and the simulation results (scattered) converge to the analytical results (solid line). The parameters used are given in Tab. \ref{tab:Analytical}.} 
\end{figure}

Fig. \ref{fig:comp_detu} shows the comparison between the analytical model and simulations of a MOC with a length between 5 cm and 1 m. As predicted by the analytical model, the brightness has a maximum for a detuning around $-\Gamma/2$. However, the brightness at a length of 5 cm is two orders of magnitude lower than the analytical model predicts. At 1 m the simulation matches the analytical result indicating that eventually the Doppler temperature and equipartition of kinetic and potential energy in the transverse direction is reached. 

\begin{figure}
    \centering
        \includegraphics[width=1.0\linewidth]{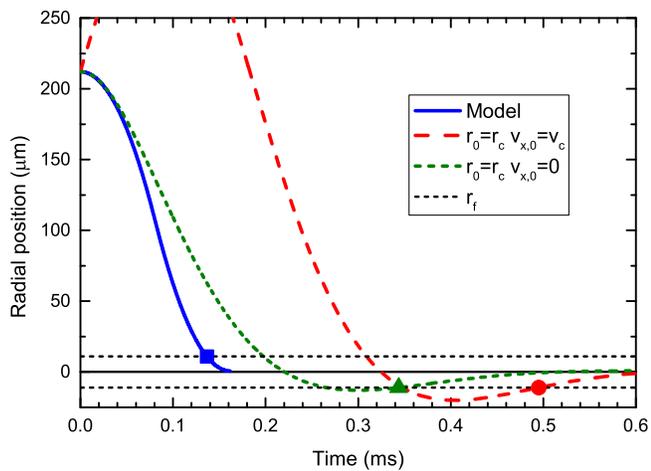}
    \caption{\label{fig:Force} Trajectories of particles with different starting conditions as a result of the actual MOC force (dashed lines) and the model force. Indicated are the times at which the particle radial position stays smaller than the final radius $r_f$.} 
\end{figure}

The difference between the model and the simulations is explained by an overestimation of the average magneto-optical force in Eq. \ref{eq:Aavg}. Numerical integration of the magneto-optical force for a particle at $r=x_c$ and $v_x=0$, as shown in figure \ref{fig:Force}, indicates that it takes the atom twice the time to get compressed into $r_f$ than expected from the model. For particles at the same starting position and a positive transverse velocity this is even larger explaining why for a detuning of $-\Gamma/2$ a MOC length of 15 cm is required to reach the brightness value as predicted by the analytical model. 

For larger detunings, the capture range, and thus the position at which the force is maximum, is larger. Particles on the edge of the aperture now experience a smaller force than the maximum and thus the average acceleration the atom will experience is lower than that given by Eq. \ref{eq:Aavg}. This requires an even longer MOC for achieving the predicted brightness.

Now the difference between the analytical model and the simulations is explained, a more accurate description of the rubidium atomic structure and the thermal source can be implemented in the simulations in order to predict the performance of the actual system.  

\subsection{Real rubidium}
In the following analysis, simulations are performed using the real atomic structure of the ${}^{85}$Rb and ${}^{87}$Rb isotope with the F=3 to F'=4 (F=2 to F'=3) cooling transitions. We only included the cooling transition and thus assume an ideal repumper is present which brings all atoms to the correct ground state and keeps them there. Tab. \ref{tab:RealRb} summarises the differences in peak-brightness and flux for the different isotopes of rubidium for the parameters in Tab. \ref{tab:Analytical}.

As a general rule, laser cooling becomes less efficient with more magnetic sublevels. This explains why the 2-level rubidium performs better than the real isotopes and why ${}^{87}$Rb performs better than the lighter isotope. Although isotopically pure ${}^{87}$Rb gives the best performance, our source will use ${}^{85}$Rb in the natural abundance as this is inexpensive. Using the parameters as provided by the analytical model, both the target brightness and flux can not be reached and thus optimization is required.
\begin{table}[t]
	\small
	\caption{\label{tab:RealRb}Performance of different rudibium isotopes.}
	\begin{tabular}{l|r|r}
		\hline
		Isotope & 10\% Brightness & 10\% flux\\
		 & A/(m${}^{2}$\,sr\,eV) & s${}^{-1}$ \\
		\hline
		2-level Rb & $7.8\times 10^5$ & $7.05\times 10^{8}$\\
		Isotopically pure ${}^{85}$Rb & $4.6\times 10^5$ & $7.05\times 10^{8}$\\
		Isotopically pure ${}^{87}$Rb & $1.5\times 10^5$ & $7.05\times 10^{8}$\\
		\hline 
		${}^{85}$Rb in natural abundance & $1.1\times 10^5$ & $5.06\times 10^{8}$\\
		${}^{87}$Rb in natural abundance & $1.3\times 10^5$ & $1.99\times 10^{8}$\\
		\hline
	\end{tabular}
\end{table}

\subsection{Collimated tube source}
In order to increase the brightness and flux of the source, a Knudsen cell with a collimation tube could be used instead of the skimmer as was used in the models. The collimation tube increases the flux density at the MOC entrance and also increases the lifetime of the source. For a heated tube with opening angle $\theta=r/L_s$ the flux is given by \cite{OlanderTube}
\begin{equation}
	\label{eq:fluxCollimated}
	\Phi_{tube} = \Phi_{tot}~8/3\cdot\theta \left( 1+ 8/3\cdot\theta \right)^{-1}.
\end{equation}
For small opening angles $\theta \ll 1$ this flux is a factor $3/(8\theta)$ higher than that of the skimmed Knudsen cell that is given by Eq. \ref{eq:fluxApertured}. Furthermore, in the skimmed source, the flux leaving the Knudsen cell is given by Eq. \ref{eq:fluxKn} but only the flux as described by Eq. \ref{eq:fluxApertured} leaves the second aperture; the difference is lost into the vacuum. The collimating tube reflects a large fraction of the flux back into the Knudsen cell reducing the loss of atoms thus increasing the lifetime of the source which is given by:
\begin{equation}
	t_{\text{life}} = \frac{N_A~M}{m~\Phi}
\end{equation} 
in which $M$ is the amount of rubidium in the source, $m$ the mass of a single rubidium atom and $N_A$ is Avogadro's number. For our set of parameters the lifetime of the skimmed source including 100 mg of Rubidium is only 70 days whereas the collimated Knudsen cell lasts for 10 years. Furthermore, the flux from this source is increased by a factor 341.

The increase in flux is impressive, but the transverse velocity distribution from the collimating tube is also much broader, resulting in a lower fraction of particles that can be captured by the MOC. To see the influence of these two competing processes, simulations were performed using the actual particle distributions for a collimating tube. The tube was implemented in the software by tracing the particles trough the tube assuming inelastic collisions with the walls following a cosine-distribution. The resulting angular distribution was verified against the theoretical prediction \cite{OlanderTube} and Monte-Carlo simulations \cite{BeijerinckTube}. 

\begin{figure}
	\includegraphics[width=1.0\linewidth]{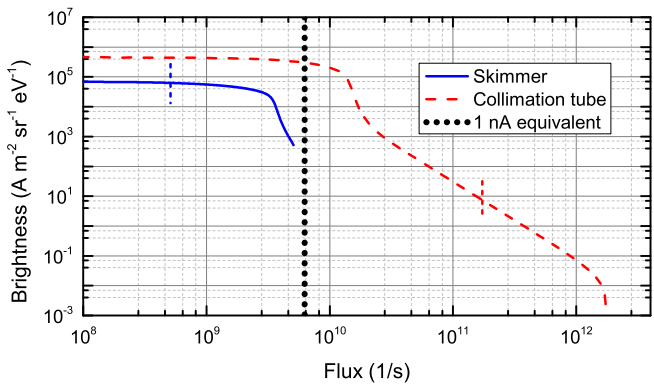}
	\caption{\label{fig:BrI_tube}Brightness-flux plot for a skimmed and a collimated Knudsen cell. Indicated are the 10\%-values for the flux. For the collimated source, both the peak brightness and the total flux are higher.}
\end{figure}

Instead of only looking at the 10\%-brightness number as was done before, this analysis is performed on a brightness-flux profile (Fig. \ref{fig:BrI_tube}). Such a profile is constructed by selecting the beam after the MOC by means of an aperture and calculating the brightness and flux of the remaining atoms. Increasing the aperture size will lead to more flux being selected and a different brightness. The flux and brightness for different aperture sizes are plotted in a single graph on the x and y-axis respectively.

Fig. \ref{fig:BrI_tube} shows how the brightness-flux profile of the skimmed and the collimated source compare for the parameters from Tab. \ref{tab:Analytical}. First, observe the much larger flux coming from the collimating tube. Then, also note the higher center brightness of the cooled and compressed beam. We credit this effect to the higher flux density in the tube. For a FIB, only the center 1 nA of current, corresponding to a flux of $6.2\times 10^{9}~\text{s}^{-1}$, is of importance, therefore we will now report on the brightness of this part of the beam. Using the collimation tube, a brightness of $3.0\times 10^5$ A/(m${}^2$\,sr\,eV) at the target flux can be achieved which is still a factor 3.3 short of the target brightness. 

\subsection{Optimisation}
So far, the parameters found by the analytical model (as given in Tab. \ref{tab:Analytical}) are used to predict the performance of the system. Here, the brightness for a flux of $6.2\times10^9~\text{s}^{-1}$ (1 nA) is optimised by variation of these parameters. 

\begin{figure}
	\includegraphics[width=1.0\linewidth]{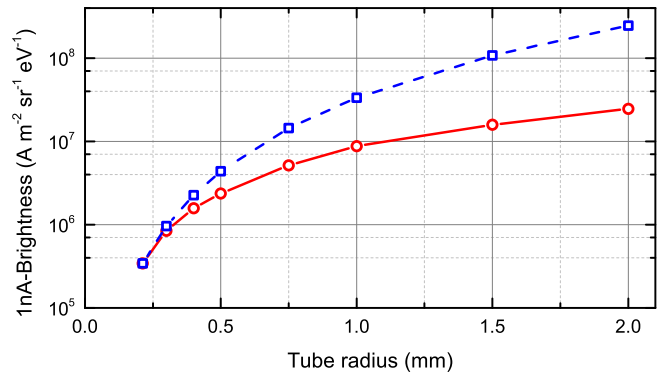}
	\caption{\label{fig:optiTubeR}The brightness (red circles) for collimation tubes with different radii but equal length. The blue squares indicate how the brightness would increase if all the additional flux could be cooled and compressed.}
\end{figure}

For the total flux scales with the third power of the collimation tube radius $r$, this is the first parameter to be considered. Introducing more atoms to the MOC will result in more particles being cooled and compressed and thus a higher brightness. Fig. \ref{fig:optiTubeR} shows the results of simulations using different tube radii, where the length of the tube has been kept constant. Increasing the tube radius does indeed result in a higher brightness (red circles), but the increase in brightness is less than what would be expected if all additional particles could be laser-cooled (blue squares). This is caused by the fact that not all the atoms at large radial position can be compressed. The increase in tube radius, and thus the flux, also decreases the source lifetime creating a trade-off between the lifetime and the brightness. Here, we chose to take the tube radius $r=1.0~\text{mm}$ as this yields a brightness 10 times higher than the target value. It decreases the lifetime only by a factor 100 to 1 year for 1 gram of rubidium in the natural mixture. 

\begin{figure}
	\includegraphics[width=1.0\linewidth]{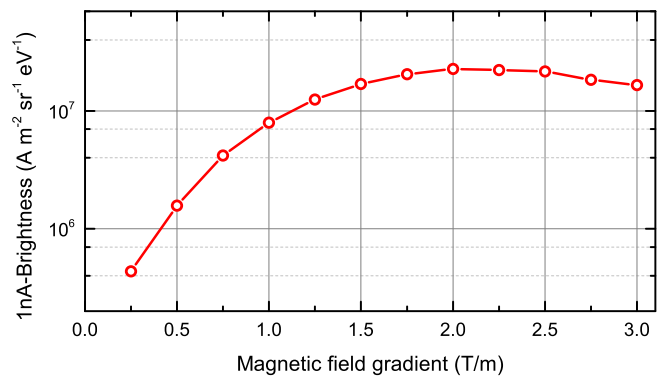}
	\caption{\label{fig:optiGrad}Brightness for different magnetic field gradients. The other parameters can be found in Tab. \ref{tab:Analytical} except for $r=1.0~\text{mm}$ and $\theta = 36~\text{mrad}$. Maximum brightness is achieved at $\nabla B=2.0~\text{T/m}$.}
\end{figure}

In the analytical model, the magnetic field gradient was chosen such that all particles could be compressed. Increasing this field results in less particles being captured but also a larger spring constant and thus a higher beam density. Simulations were performed to check at which scale of the magnetic field gradient this trade-off is important. As is shown in Fig. \ref{fig:optiGrad} an optimum indeed exists at a gradient strength of 2.0 T/m which improves the brightness by a factor 2.3.

\begin{figure}
	\includegraphics[width=1.0\linewidth]{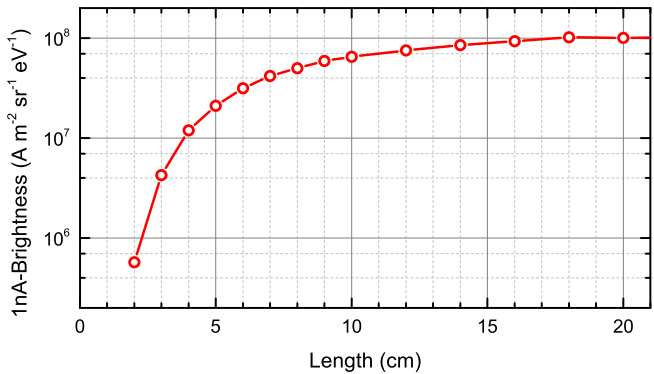}
	\caption{\label{fig:optiLength}Brightness for different MOC lengths. The other parameters can be found in Tab. \ref{tab:Analytical} except for $r=1.0~\text{mm}$, $\theta = 36~\text{mrad}$ and $\nabla B=2.0~\text{T/m}$. Increasing the MOC length does improve the brightness but not to great extent.}
\end{figure}

The final parameter to check is the MOC length $L_c$. Fig. \ref{fig:optiLength} shows that the compact 5 cm long MOC performs quite well: making it twice as long increases the brightness by a factor 2.5 whereas making it twice as short decreases it by a factor 7. At 5 cm the brightness at 1 nA reads $2.1\times 10^7$ A/(m${}^2$\,sr\,eV) which is more than a factor 10 higher than the target. This brightness value corresponds to a flux density at the MOC end of $4\times 10^{19}$ m${}^{-2}$\,s${}^{-1}$ and a transverse temperature of 2 mK.

The performance may be improved even further by increasing the saturation parameter or source temperature. Although increasing the saturation parameter raises the Doppler temperature, it also leads to more particles being captured. As the simulation program does not include cross-saturation effects like stimulated emission by the opposite laser beam, the high saturation parameter regime was not investigated. The same holds for the higher source temperature: an increase leads to more inter-particle collisions in the collimation tube, making the transverse velocity distribution broader than that of the collision-free model that is used in the simulations. We expect the effect of this broadening to be less important than the increase in flux from the higher temperature. The brightness therefore still increases, but less than what would be expected from the higher flux.   

\section{Conclusion}
In this paper we have discussed the possibility of creating a high quality atomic beam of rubidium as a a precursor for a high brightness ionic beam for use in a FIB. First, the proposed setup consisting of a collimated Knudsen cell and a 2D MOC was introduced. An analytical model was set up to predict the performance of the system and to gain insight in the relevant parameters. This model was used to verify Monte Carlo simulations on the laser cooling process. The simulation program was extended to include the atomic structure of the ${}^{85}$Rb and ${}^{87}$Rb isotopes and the initial distribution distributions for a the collimated tube source. Optimization of the relevant parameters lead to a predicted brightness of $2.1\times 10^7$ A/(m${}^2$\,sr\,eV) at a flux of $6.2 \times 10^9$ s${}^{-1}$ (equivalent to a current of 1 nA) which is more than an order of magnitude higher than the state of the art LMIS. In additional work \cite{Haaf2014} simulations are performed on a realistic ionization and accelerator structure and a focusing column. These simulations show that sub-nanometer spot sizes can be achieved at a current of 1 pA.

\section{Acknowledgements}
This research is supported by the Dutch Technology Foundation STW, applied science division of the ``Nederlandse Organisatie voor Wetenschappelijk Onderzoek (NWO)'', FEI Company, Pulsar Physics and Coherent Inc.

\bibliography{LaserCooling}

\end{document}